\documentclass[12pt,a4paper]{amsart}
\usepackage{amsmath,amssymb,a4wide,amsthm,ifthen,hyperref,xcolor,enumerate,enumitem}
\usepackage[utf8]{inputenc}

\usepackage{graphicx}
\usepackage{url}
\usepackage{ulem}
\usepackage{natbib}
\usepackage{flafter} 
\usepackage{multirow}
\usepackage{float}
\usepackage{tabularx,ragged2e}
\usepackage{booktabs,ulem}
\usepackage{makecell}
\newcolumntype{Y}{>{\centering\arraybackslash}X}

\theoremstyle{remark}

\newcommand{\by}{\mathbf{y}}
\newcommand{\bX}{\mathbf{X}}
\newcommand{\bSigma}{\boldsymbol{\Sigma}}

\newcommand{\bbeta}{\boldsymbol{\beta}}
\newcommand{\bepsilon}{\boldsymbol{\epsilon}}

\newcommand{\norm}[1]{\left\lVert#1\right\rVert}
\def\argmin{\mathop{\rm Argmin}\nolimits}

\begin{document}

\title[A variable selection approach for highly correlated predictors]{A variable selection approach for highly correlated predictors in high-dimensional genomic data}

\date{}

\author{Wencan Zhu}
\address{UMR MIA-Paris, AgroParisTech, INRAE, Universit\'e Paris-Saclay, 75005, Paris, France}
\email{wencan.zhu@agroparistech.fr}
\author{Céline Lévy-Leduc}
\address{UMR MIA-Paris, AgroParisTech, INRAE, Universit\'e Paris-Saclay, 75005, Paris, France}
\email{celine.levy-leduc@agroparistech.fr}
\author{Nils Ternès}
\address{Biostatistics and Programming department, Sanofi R\&D, 91380 Chilly Mazarin, France}
\email{nils.ternes@sanofi.com}

\keywords{variable selection; highly correlated predictors; genomic data}

\maketitle

\begin{abstract}
 In genomic studies, identifying biomarkers associated with a variable of interest is a major concern in biomedical research. Regularized approaches are classically used to perform variable selection in high-dimensional linear models.
  However, these methods can fail in highly correlated settings.
  We propose a novel variable selection approach called WLasso, taking these correlations into account. It consists in rewriting the initial high-dimensional linear model to remove the correlation between the biomarkers (predictors) and in applying the
  generalized Lasso criterion. The performance of WLasso is assessed using synthetic data in several scenarios and compared with recent alternative approaches. The results show that when the biomarkers are highly correlated, WLasso outperforms the other approaches in sparse high-dimensional frameworks. The method is also successfully illustrated on publicly available gene expression data in breast cancer. 
 Our method is implemented in the \texttt{WLasso} R package which
  is available from the Comprehensive R Archive Network.
\end{abstract}

\section{Introduction}

The identification of prognostic genomic biomarkers (i.e. biomarkers associated with a variable of interest, for example a clinical endpoint in clinical trials) has become a major concern for the biomedical research field. Indeed, prognostic biomarkers may help to anticipate the prognosis of individual patients and may
also be useful to understand a disease at a molecular level and possibly guide for the development of new treatment strategies (\cite{Kalia:2015}).

To this end, statistical variable selection approaches are widely used to identify a subset of biomarkers in high-dimensional settings where the number of biomarkers $p$ is much larger than the sample size $n$.
Several reviews focused on this topic (\cite{Saeys07} and \cite{Reviewpractic} for example). Commonly used techniques include hypothesis-based test: t-test (\cite{Ttest}), wrapper approaches (\cite{Saeys07}): forward, backward selection, and penalized approaches: Lasso (\cite{Lasso}), Elastic-net (\cite{Zou:2015}), SCAD (\cite{SCAD}) among others. 
Hypothesis tests are limited to independently consider associations for each biomarker thus neglecting potential relationships between them.
Wrapper approaches often show high risk of overfitting and are computationally expensive for high-dimensional data (\cite{Knecht:2005}).
More efforts have been devoted to penalized methods, given the attractive feature of automatically performing variable selection
and coefficient estimation simultaneously (\cite{Fan:2006}). We shall thus focus on this type of approaches in the following.

Let us consider the following linear regression model: 
\begin{equation}
\label{lm}
    \by=\bX\bbeta+\bepsilon.
\end{equation}
where $\by=(y_{1}, \ldots, y_{n})^{T}$ is the variable to explain (clinical endpoint), $\bX=(\bX_{1}, \ldots, \bX_{p})$ is the design matrix containing the
expression of biomarkers such that the correlation matrix of its columns is  $\boldsymbol{\Sigma}$, $\bbeta=(\beta_{1}, \ldots, \beta_{p})^{T}$
is a sparse vector to estimate, namely with \textcolor{black}{a majority of} null coefficients, and $\bepsilon$ is the error term.
The Lasso penalty is a well-known approach to estimate $\bbeta$ with a sparsity enforcing constraint. It consists in minimizing the following
penalized least-squares criterion (\cite{Lasso}):
\begin{equation}
L_{\lambda}(\bbeta)=\norm{\by-\bX\bbeta}_{2}^{2}+\lambda\norm{\bbeta}_{1},
\end{equation}
where $\norm{\cdot}_{2}$ is the Euclidean norm and $\norm{\bbeta}_{1}=\sum_{k=1}^p |\beta_k|$.
However, the Lasso has several drawbacks in \textcolor{black}{highly} correlated settings (\cite{Zou:2015}) such as the violation of the Irrepresentable Condition (IC)
defined in \cite{Zhao:2006}. The authors of this article prove that this condition is necessary and sufficient to recover the support of $\bbeta$,
namely to retrieve the null and non null components
in the vector $\bbeta$  and thus to provide a sign consistent estimator. This condition is defined as follows.
Let $S=\{j,\;\beta_{j}\neq0\}$ be the set of active variables,
$S^{c}$ the set of non-active variables and $\bX_A$ the submatrix of $\bX$ containing only the indices of columns which are in the set $A$.
Then, the design matrix $\bX$ satisfies the IC if, for some constant $\eta \in (0, 1]$,
\begin{equation}
\label{IC equation}
\left|\left(\bX_{S^{c}}^{T}\bX_{S}(\bX_{S}^{T}\bX_{S})^{-1}\textrm{sign}(\bbeta_{S})\right)_j\right| \leq 1-\eta, \textrm{ for all } j,
\end{equation}
where $\textrm{sign}(x)=1$, if $x>0$, -1 if $x<0$ and 0 if $x=0$. Intuitively, this condition means that the correlation between the active and non active
explanatory variables is smaller that the correlation between the active explanatory variables. Hence,
this condition is most likely to be violated when the correlations between non active and active variables are large.
In high-dimensional genomic data, this condition is difficult to guarantee as the correlation between biomarkers is usually high (\cite{Michalopoulos:2012}).
\cite{Wang:2018} tested the irrepresentable condition on several publicly available genomic data and highlighted that the condition is violated in almost all the
\textcolor{black}{datasets} investigated.

Methods have been proposed to deal with the issue of high correlations between the biomarkers.
Preconditioning the Lasso is one of them. It consists in transforming the given data $\bX$ and $\by$ before applying the Lasso criterion. For example, \cite{jia:2015} and \cite{holp:2016} proposed to left multiply $\bX$, $\by$ and thus $\bepsilon$ in Model (\ref{lm}) by \textcolor{black}{specific matrices to remove the correlations between the columns of $\bX$}. 
\textcolor{black}{A major drawback of} the latter, called HOLP (High dimensional Ordinary Least squares Projection), is that
the preconditioning step may \textcolor{black}{increase the variance of the error term} and thus may alter the variable selection performance. Another recently published method
\textcolor{black}{named} Precision Lasso (\cite{Wang:2018}) proposes to handle the correlation issue by assigning similar weights to correlated variables.
This approach revealed better performance than the other methods when the biomarkers were highly correlated. However, it failed in more favorable
cases when \textcolor{black}{the biomarkers} are not correlated.

In this paper, we propose an alternative and novel approach, called \textcolor{black}{Whitening Lasso} (WLasso), \textcolor{black}{to take into account the
  correlations that may exist between the predictors (biomarkers).} 
Our method proposes to transform Model (\ref{lm}) in order to remove the correlations existing between the columns of $\bX$ and thus to ``whiten'' them and
make the IC valid but without changing the error term $\bepsilon$. This prevents us from noise inflation, see (\ref{lm_modified}).
Then, the variable (biomarker) selection is performed thanks to the generalized Lasso criterion devised by \cite{tibshirani:2011}.
The full details of our method are provided in Section \ref{sec:methods}. An extensive simulation study is presented in Section \ref{sec:numexp} to assess
\textcolor{black}{the selection performance of our approach and to compare it} to other methods in different settings.
WLasso is also applied to a publicly available dataset in breast cancer in Section \ref{sec:real}. Finally, we discuss our
findings and give concluding remarks in Section \ref{sec:conclusion}.


\section{Methods}\label{sec:methods}

In this section, we propose a novel variable selection approach called \textsf{WLasso} (Whitening Lasso) which consists in removing the
  correlations existing between the biomarkers (columns of $\bX$) and in applying the generalized Lasso criterion proposed by \cite{tibshirani:2011}
for variable selection purpose.

\subsection{Model Transformation}\label{sec:model_transform}

Inspired by the literature on preconditioning, we propose to rewrite Model (\ref{lm}) in order to remove the correlation existing between the columns of $\bX$.
More precisely, let $\boldsymbol{\Sigma}^{-1/2}:=\boldsymbol{U}\boldsymbol{D}^{-1/2}\boldsymbol{U}^{T}$ where $\boldsymbol{U}$ and $\boldsymbol{D}$ are the matrices involved in the spectral decomposition
of  the symmetric matrix $\boldsymbol{\Sigma}$ given by: $\boldsymbol{\Sigma}=\boldsymbol{U}\boldsymbol{D}\boldsymbol{U}^{T}$. We then denote $\widetilde{\bX}=\bX\boldsymbol{\Sigma}^{-1/2}$. Therefore, (\ref{lm}) can be rewritten as follows:
\begin{equation}\label{lm_modified}
    \by=\widetilde{\bX}\widetilde{\bbeta}+\bepsilon,
\end{equation}
where $\widetilde{\bbeta}=\boldsymbol{\Sigma}^{1/2}\bbeta:=\boldsymbol{U}\boldsymbol{D}^{1/2}\boldsymbol{U}^{T}\bbeta$. 
With such a transformation, since the $n$ rows $\boldsymbol{x}_1,\dots,\boldsymbol{x}_n$ of $\bX$ are assumed to be independent Gaussian random vectors with a covariance matrix equal to $\boldsymbol{\Sigma}$, the covariance matrix of the rows of $\widetilde{\bX}$ is equal to identity and the columns of 
$\widetilde{\bX}$ are thus uncorrelated. The advantage of such a transformation with respect to the preconditioning approach proposed by \cite{holp:2016} is that the error term $\bepsilon$ is not modified thus avoiding an increase of the noise which can overwhelm the benefits of a well conditioned design matrix.

\begin{figure}
\centering
\includegraphics[scale=0.25]{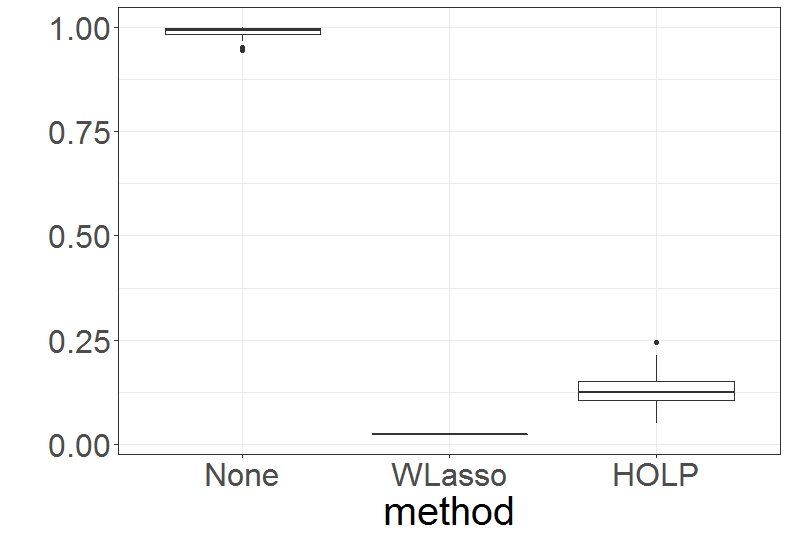}
  \caption{Proportion of components $j$ such that (\ref{IC equation}) is violated. These results were obtained from 100 replications.\label{fig:IC}}
\end{figure}

To illustrate the benefits of our methodology,  observations $\by$ were generated according to Model (\ref{lm}) with $p=500$, $n=50$, $\bbeta$ having 10 non null components 
which are equal to 2 and with $\boldsymbol{\Sigma}$ defined by
\begin{equation}
     \label{eq:SPAC}
     \bSigma=
       \begin{bmatrix}
         \bSigma_{11} &  \bSigma_{12} \\
         \bSigma_{12}^{T} &  \bSigma_{22}
       \end{bmatrix}
      \end{equation} 
where $\bSigma_{11}$ is the correlation matrix of active variables with off-diagonal entries equal to $\alpha_1$, $\bSigma_{22}$ is the one of non active variables with off-diagonal entries equal to 
$\alpha_3$ and $\bSigma_{12}$ is the correlation matrix between active and non active variables with entries equal to $\alpha_2$. In the case where $(\alpha_1,\alpha_2,\alpha_3)=(0.3, 0.5, 0.7)$,
Figure \ref{fig:IC} displays the percentage of components $j$ for which the Irrepresentable Condition (\ref{IC equation}) is not satisfied
from 100 replications. We can see from this figure that our approach 
(\textsf{WLasso}) dramatically improves the number of indices $j$ for which the IC condition is satisfied. The results are even better than those obtained by the transformation proposed by \textsf{HOLP} (\cite{holp:2016}).

The following illustrations of Section \ref{sec:methods} are obtained from observations $\by$ generated according to the previous scenario.

\subsection{Estimation of $\widetilde{\bbeta}$}\label{sec:beta_tilde}
In order to estimate $\widetilde{\bbeta}$ with a sparsity enforcing constraint on $\bbeta$, we use the generalized Lasso criterion proposed by \cite{tibshirani:2011} which consists
in minimizing the following criterion with respect to $\bbeta$:
\begin{equation*}
\norm{\by-\bX\bbeta}_{2}^{2}+\lambda\norm{\mathbf{D}\bbeta}_{1}, 
\end{equation*}
where $\mathbf{D}$ is a specific matrix.
Note that this criterion boils down to the classical Lasso criterion if $\mathbf{D}$ is the identity matrix.
In Model (\ref{lm_modified}), we thus propose to minimize the following criterion with respect to $\widetilde{\bbeta}$:  
\begin{equation}\label{genlasso}
L_{\lambda}^{\textrm{gen}}(\widetilde{\bbeta})=\norm{\by-\widetilde{\bX}\widetilde{\bbeta}}_{2}^{2}+\lambda\norm{\boldsymbol{\Sigma}^{-1/2}\widetilde{\bbeta}}_{1},
\end{equation}
which guarantees a sparsity enforcing constraint on $\bbeta$ thanks to the $\ell_1$ penalty. We thus obtain
$$
\widehat{\widetilde{\bbeta}}_{0}(\lambda)={\argmin}_{\widetilde{\bbeta}}\; L_{\lambda}^{\textrm{gen}}(\widetilde{\bbeta}).
$$
To estimate $\widetilde{\bbeta}$, we will not directly use $\widehat{\widetilde{\bbeta}}_{0}(\lambda)$ but the following modified estimator which can be seen as a thresholding of the components of $\widehat{\widetilde{\bbeta}}_{0}(\lambda)$.
For $K$ in $\{1,\ldots,p\}$, let $\textrm{Top}_K$ be the set of indices corresponding
to the $K$ largest values of the components of $|\widehat{\widetilde{\bbeta}}_{0}|$, then
the estimator of $\widetilde{\bbeta}$ is $\widehat{\widetilde{\bbeta}}=(\widehat{\widetilde{\bbeta}}_j^{(\widehat{K})})$
where $\widehat{\widetilde{\bbeta}}_j^{(K)}$ is defined by:
\begin{equation}\label{eq:beta_tilde_thresh}
\widehat{\widetilde{\bbeta}}_{j}^{(K)}(\lambda)=
\begin{cases}
  \widehat{\widetilde{\bbeta}}_{0j}(\lambda), & j \in \textrm{Top}_K \\
 \textrm{$K$th largest value of } |\widehat{\widetilde{\bbeta}}_{0j}| , & j \not\in \textrm{Top}_K.
\end{cases}
\end{equation}
The choice of $\widehat{K}$ is explained in Section \ref{subsec:choice}.

\begin{figure}
  \centering
  \includegraphics[scale=0.25,trim={1.2cm 2.2cm 0 0},clip]{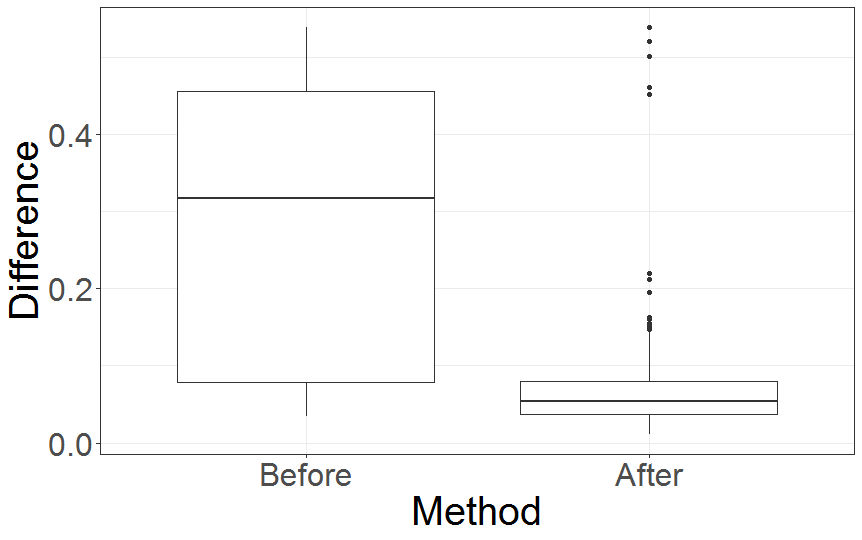}
\centering
\caption{Boxplots of the average of $(|\widehat{\widetilde{\bbeta}}_{0j}(\lambda)-\widetilde{\bbeta}_j(\lambda)|)_{1\leq j\leq p}$ (left) and
  $(|\widehat{\widetilde{\bbeta}}_j^{(\widehat{K})}(\lambda)-\widetilde{\bbeta}_j(\lambda)|)_{1\leq j\leq p}$ for all $\lambda$ (right)
  obtained from 100 replications.
  \label{fig:beta_tilde}}
\end{figure}

Figure \ref{fig:beta_tilde} displays the average of $(|\widehat{\widetilde{\bbeta}}_j^{(\widehat{K})}(\lambda)-\widetilde{\bbeta}_j(\lambda)|)_{1\leq j\leq p}$
for all the values of $\lambda$ that are considered. 
We can see from this figure that the thresholding improves the estimation of $\widetilde{\bbeta}$.

\subsection{Estimation of $\bbeta$}\label{sec:estim_beta}
As previously, to estimate $\bbeta$, we will first consider
  $\widehat{\bbeta}_0=\boldsymbol{\Sigma}^{-1/2}\widehat{\widetilde{\bbeta}}$
  and then apply a thresholding strategy. Thus, we propose to estimate $\bbeta$ by $\widehat{\bbeta}=(\widehat{\bbeta}_j^{(\widehat{M})})_{1\leq j\leq p}$
  where $\widehat{\bbeta}_j^{(M)}$ is defined by: 
\begin{equation}\label{eq:beta_hat}
\widehat{\bbeta}_{j}^{(M)}(\lambda)=
\begin{cases}
  \widehat{\bbeta}_{0j}(\lambda), & j \in \textrm{Top}_M \\
  0 , & j \not\in \textrm{Top}_M.
\end{cases}
\end{equation}
The choice of $\widehat{M}$ is explained in Section \ref{subsec:choice}.

\begin{figure}
  \hspace{-5mm}
  \includegraphics[scale=0.28,trim={3cm 0 2cm 0},clip]{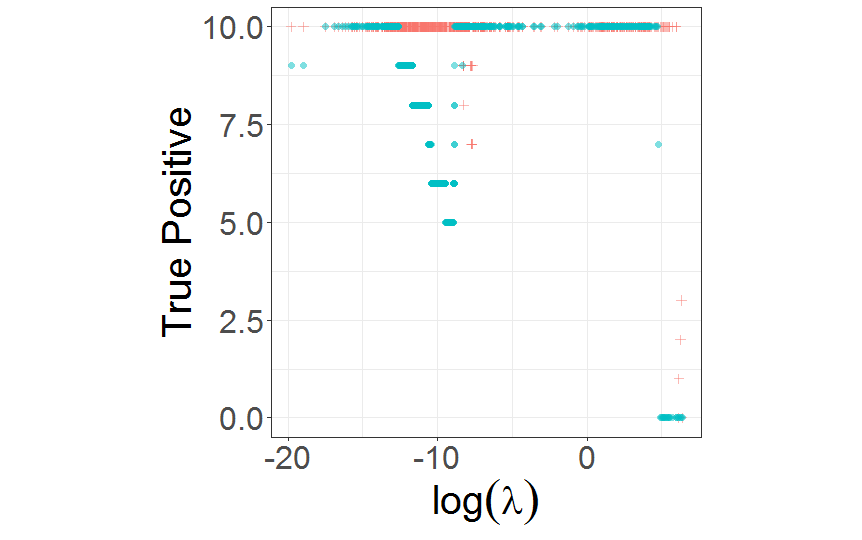}
  \hspace{-6mm}
  \includegraphics[scale=0.28,trim={3cm 0 2cm 0},clip]{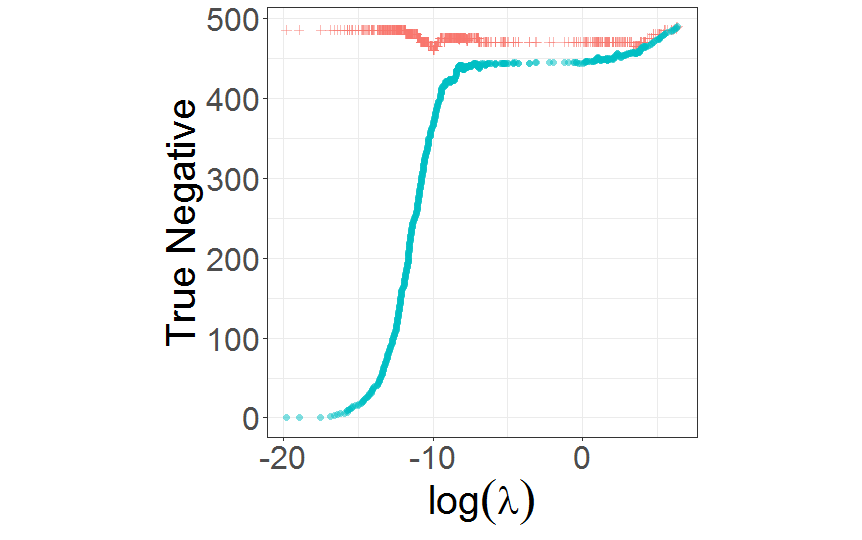}
  \caption{Number of True Positive and True Negative for $\widehat{\bbeta}$ in red and $\widehat{\bbeta}_0$ in blue
    for a given vector of observations $\by$.\label{fig:beta_final}}
\end{figure}

As we can see from Figure \ref{fig:beta_final}, more true non null (active) components of $\bbeta$ (true positive)
and more true null (non active) components of $\bbeta$ (true negative) can be retrieved with $\widehat{\bbeta}$ than with $\widehat{\bbeta}_0$.

\subsection{Choice of the parameters}\label{subsec:choice}

To choose the parameters $K$ and $M$ in (\ref{eq:beta_tilde_thresh}) and (\ref{eq:beta_hat}) for each $\lambda$,
we use a strategy based on the Mean Squared Error (MSE). We shall first explain the strategy that we used for choosing $\widehat{K}$.
Let
$$
\widetilde{\textrm{MSE}}_{K}(\lambda)=\|\by-\widetilde{\bX}\widehat{\widetilde{\bbeta}}^{(K)}(\lambda)\|_2^2,
$$
where $\by$, $\widetilde{\bX}$ and $\widehat{\widetilde{\bbeta}}^{(K)}(\lambda)$ are defined in (\ref{lm}), (\ref{lm_modified}) and (\ref{eq:beta_tilde_thresh}), respectively and
$$
\widehat{K}(\lambda)=\argmin \left\{K\geq 1 \textrm{ s.t. } \frac{\widetilde{\textrm{MSE}}_{K+1}(\lambda)}{\widetilde{\textrm{MSE}}_{K}(\lambda)}\geq\gamma\right\},\textrm{ where } \gamma\in (0,1).
$$
Large values of $\gamma$ will lead to large values of $\widehat{K}(\lambda)$ and thus to a weak thresholding of the estimator of $\widetilde{\bbeta}$.
In practice, as it is shown in Section \ref{sec:numexp}, taking $\gamma$ in (0.9,0.99) provides satisfactory and almost similar results.

\begin{figure}
  \includegraphics[scale=0.28,trim={4cm 0 3cm 0},clip]{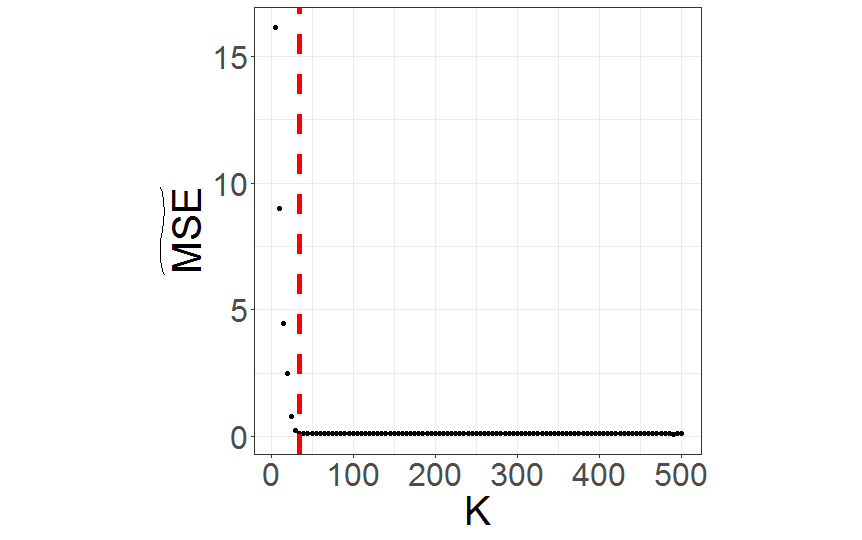}
  \hspace{-4mm}
  \includegraphics[scale=0.28,trim={4cm 0 3cm 0},clip]{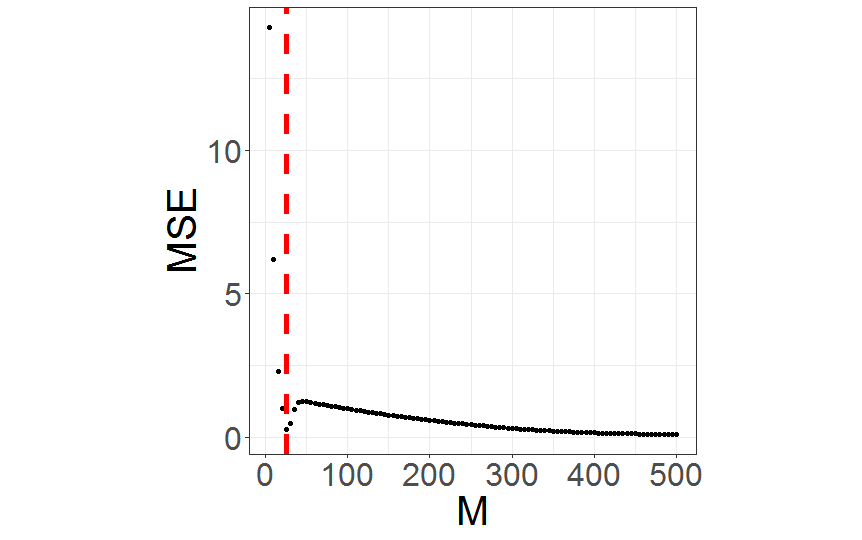}
  \caption{$\widetilde{\textrm{MSE}}_{K}(\lambda)$ (left) and $\textrm{MSE}_{M}(\lambda)$ (right) for $\lambda$ chosen thanks to the strategy
    explained in Section \ref{subsec:lambda} for a given vector of observations $\by$. The vertical dotted lines correspond to $\widehat{K}(\lambda)$ and $\widehat{M}(\lambda)$, respectively.\label{fig:MSE}}
\end{figure}

For the choice of $\widehat{M}(\lambda)$, we use the same procedure except that $\widetilde{\textrm{MSE}}_{K}(\lambda)$ is replaced
by
\begin{equation}\label{eq:MSE}
\textrm{MSE}_{M}(\lambda)=\|\by-\bX\widehat{\bbeta}^{(M)}(\lambda)\|_2^2,
\end{equation}
where $\by$, $\bX$ and $\widehat{\bbeta}^{(M)}(\lambda)$ are defined in (\ref{lm}) and (\ref{eq:beta_hat}), respectively.
Both criteria are displayed in Figure \ref{fig:MSE} for a value of $\lambda$ which is chosen according to
the strategy explained in Section \ref{subsec:lambda}.

\subsection{Estimation of $\bSigma$}\label{sec:estim_Sigma}
Since the matrix $\bSigma$ is unknown in practice, it has to be estimated. In the particular situation where $\bSigma$ has the block
  structure described in (\ref{eq:SPAC}), we propose the following strategy. Firstly, we compute the empirical correlation matrix as follows.
Let $\boldsymbol{S}$ be the sample $p\times p$ covariance matrix defined by
$$
\boldsymbol{S}=\frac{1}{n-1}\sum_{i=1}^n  \left(\boldsymbol{x}_{i}-\overline{\boldsymbol{x}}\right)
\left(\boldsymbol{x}_{i}-\overline{\boldsymbol{x}}\right)',\quad\textrm{with } \overline{\boldsymbol{x}}=\frac{1}{n}\sum_{i=1}^n \boldsymbol{x}_{i},
$$
where $\boldsymbol{x}_{i}$ denotes the $i$th row of $\boldsymbol{X}$ defined in (\ref{lm}). The corresponding $p\times p$ sample correlation matrix $\boldsymbol{R}=(R_{i,j})$ is defined by:
\begin{equation}\label{eq:mat_corr}
\boldsymbol{R}_{i,j}=\frac{S_{i,j}}{\sigma_{i}\sigma_{j}},\; \forall 1\leq i,j\leq p,
\end{equation}
where
$$
\sigma_{i}^2=\frac{1}{n-1}\sum_{\ell=1}^n (X_{\ell,i}-\overline{X}_i)^2,\quad\textrm{with } \overline{X}_i=\frac{1}{n}\sum_{\ell=1}^n X_{\ell,i},\; \forall 1\leq i\leq p.
$$
Secondly, the two groups (or clusters) of active and non active biomarkers are obtained by using a hierarchical clustering with
the complete agglomeration method.
Thirdly, the entries of $\widehat{\bSigma}$ are computed by averaging the values of $\boldsymbol{R}$ within the groups.
More precisely, let  $\rho_{i,j}$ denote the value of the entries in the 
block having its rows corresponding to Cluster $i$ and its columns to Cluster $j$.
Then, for a given clustering $C$:
\begin{equation}\label{eq:clust2Sig}
\rho_{i,j}= \left\{ \begin{tabular}{lcl}
$\frac{1}{\#C(i)\#C(j)}\displaystyle\sum_{k \in C(i), \ell \in C(j)}R_{k,\ell}$, & if $ C(i) \neq C(j)$ \\
  & \\
 $\frac{1}{\#C(i)(\#C(i)-1)}\displaystyle\sum_{k \in C(i), \ell \in C(i), k\neq \ell}R_{k,\ell}$, & if $ C(i) = C(j)$
 \end{tabular} \right.,
 \end{equation}
 where $C(i)$ denotes the cluster $i$, $\#C(i)$ denotes the number of elements in the cluster $C(i)$ and $R_{k,\ell}$ is the $(k,\ell)$ entry  of the matrix $\boldsymbol{R}$ 
defined in (\ref{eq:mat_corr}).

\begin{figure}
  \begin{center}
  \includegraphics[scale=0.22,trim={0cm 0 0cm 0},clip]{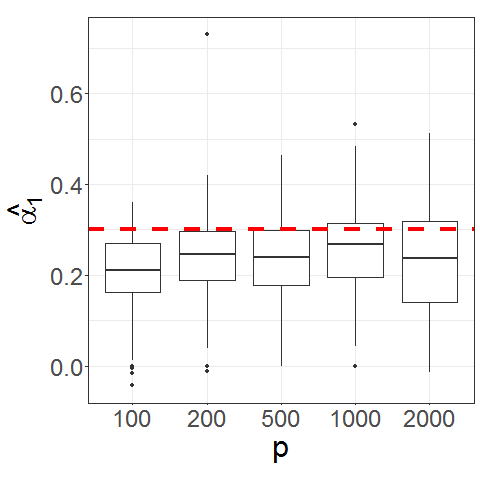}
  \includegraphics[scale=0.22,trim={0cm 0 0cm 0},clip]{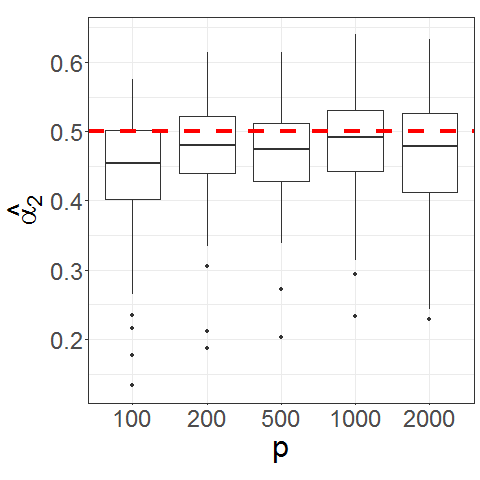}
  \includegraphics[scale=0.22,trim={0cm 0cm 0cm 0},clip]{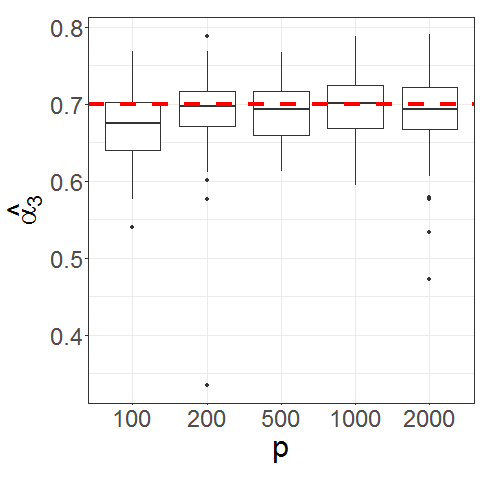}
\end{center}
\caption{Estimation of the parameters $(\alpha_1,\alpha_2,\alpha_3)=(0.3,0.5,0.7)$. The horizontal dotted lines correspond to the true values of the
    parameters. These results are obtained from 100 replications for each value of $p$.\label{fig:est_alpha_sigma}}
\end{figure}

We illustrate the performance of our method in Figure \ref{fig:est_alpha_sigma}
in the case where $\bSigma$ has the structure (\ref{eq:SPAC}) with $(\alpha_1,\alpha_2,\alpha_3)=(0.3,0.5,0.7)$.
%
%
We can see from this figure that the proposed methodology for estimating the correlation coefficients within the blocks of
$\widehat{\bSigma}$ is very efficient.

\subsection{Summary of the WLasso method}\label{sec:summary}

The WLasso method can be summarized as follows:
\begin{itemize}
  \item \underline{\textsf{First step}}: Estimation of the matrix $\bSigma$ by $\widehat{\bSigma}$, see Section \ref{sec:estim_Sigma}.
    \item \underline{\textsf{Second step}}: Transformation of Model (\ref{lm}) into Model (\ref{lm_modified}) to remove the correlation existing between the columns of $\bX$, see Section \ref{sec:model_transform} where $\bSigma$ is replaced by $\widehat{\bSigma}$.
    \item \underline{\textsf{Third step}}: Estimation of $\widetilde{\bbeta}$ defined in (\ref{lm_modified}), see Section \ref{sec:beta_tilde}.
    \item \underline{\textsf{Fourth step}}: Estimation of $\bbeta$ defined in (\ref{lm}), see Section \ref{sec:estim_beta} and identification of its null and
      non null components.
    \end{itemize}


\section{Numerical experiments}\label{sec:numexp}

We performed numerical experiments to assess the performance of the WLasso and to compare it with other recent approaches.

All simulated datasets were generated from Model (\ref{lm}) in which the number of predictors (biomarkers) $p$ is equal to 100, 200, 500, 1000 or 2000
and the sample size $n$ is equal to 50 or 100.
We randomly chose 10 non null coefficients  among the $p$ coefficients of $\bbeta$ which correspond to the active biomarkers, thus considering different
sparsity levels. The value $b$ of the non null coefficients
is equal to either 0.5 or 1 to consider different signal-to-noise ratios.  

Regarding the correlation matrix $\bSigma$ which contains the correlation values between the biomarkers, namely the correlations between the columns
of the design matrix $\bX$, several structures were considered:
\begin{itemize} 
\item Block-wise correlation structure defined in (\ref{eq:SPAC}) with parameters $(\alpha_{1} , \alpha_{2} , \alpha_{3})=(0.3, 0.5, 0.7)$ and $(0.5, 0.7, 0.9)$;
\item Independent setting where $\bSigma$ is the identity matrix.
\end{itemize}
The results that are presented hereafter are obtained from 100 replications.

\subsection{Estimation of $\boldsymbol{\Sigma}$}

\begin{figure}[!h]
  \begin{center}
    \includegraphics[scale=0.5]{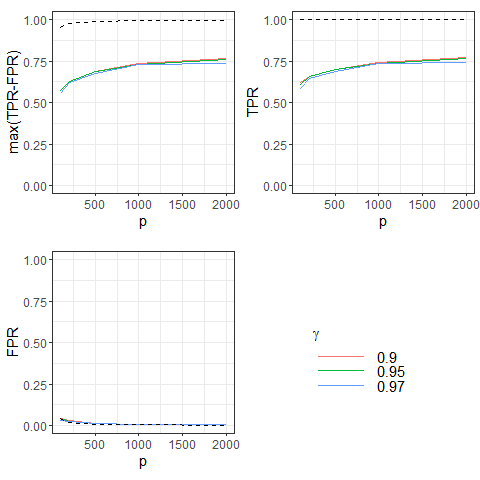}
  \end{center}
  \caption{max(TPR-FPR) and the corresponding True Positive Rate (TPR) and False Positive Rate (FPR) for $(\alpha_{1}, \alpha_{2}, \alpha_{3})=(0.3, 0.5, 0.7)$, $b=0.5$ and $n=50$. Dotted line: $\bSigma$, solid line: $\widehat{\bSigma}$. \label{fig:Oracle_compare}}
\end{figure}

To evaluate the impact of the estimation of $\bSigma$, simulations were performed to compare the performance
of WLasso when $\bSigma$ is known and when it is estimated. The results are displayed in
Figure \ref{fig:Oracle_compare} for several values of $\gamma$ (0.9, 0.95, 0.97) which is a parameter appearing
in Section \ref{subsec:choice}. In the top left part of this figure the largest difference between the 
True Positive Rate (TPR) and False Positive Rate (FPR) is displayed for several values of $p$
and for $n=50$. In the top right and bottom parts of the figure, the corresponding
TPR and FPR are displayed, respectively. We can see from this figure that for the value of $\lambda$ maximizing the
difference between TPR and FPR and for all the values of $\gamma$, all the active variables are properly retrieved 
without selecting non active variables when $\bSigma$ is known. In the case where $\bSigma$ is estimated by using the
approach described in Section \ref{sec:estim_Sigma}, 75\% of the active variables are recovered and less than 1\% of non active variables are wrongly estimated as active variables. Note that the results displayed in Figure \ref{fig:Oracle_compare} are obtained when $(b,n)=(0.5,50)$
but we obtained similar conclusions for $(b,n)=(1,50)$, $(b,n)=(0.5,100)$ and $(1,100)$.


\subsection{Choice of $\lambda$}\label{subsec:lambda}

For tuning the parameter $\lambda$ involved in our methodology, we propose choosing the value which minimizes
$\textrm{MSE}_{\widehat{M}(\lambda)}(\lambda)$ defined in (\ref{eq:MSE}).
In Figure \ref{fig:Opt_compare_suppl} of the Supplementary material, we compare the performance of our approach with this choice of $\lambda$ (solid line)
to the optimal one obtained when $\lambda$ is chosen to yield the largest difference between the TPR and the FPR (dotted line).
We can observe from this figure that, for the different values of $\gamma$, the TPR is a little bit smaller when $\lambda$ is estimated but that
the FPR is quite similar.  Similar results were obtained for $(b,n)=(1,50)$, $(b,n)=(0.5,100)$ and $(1,100)$. However,
for larger $b$ or $n$, the difference between the performance obtained by the optimal choice of $\lambda$ and by our choice of $\lambda$ is smaller.


\subsection{Comparison with other methods}

In this section, we compare our methodology with other approaches: the classical Lasso described in \cite{Lasso}
and two recently proposed methods aiming at handling
the correlations between the columns of the design matrix $\bX$: HOLP  and Precision Lasso proposed by \cite{holp:2016} and \cite{Wang:2018}, respectively.
This comparison is performed by computing the TPR and FPR of these approaches for different values of the parameters involved in each of them.


The grid of $\lambda$ for the classical Lasso and for our approach is provided by the \texttt{glmnet} and \texttt{genlasso} R packages,
  respectively. Concerning the Precision Lasso, we found for each value of $n$ and $p$
the $\lambda_{\textrm{min}}$ and $\lambda_{\textrm{max}}$ leading
to $p$ non null estimated coefficients and $p$ null estimated coefficients, respectively.
Then, we chose 100 values of $\lambda$ uniformly distributed in the interval $[\lambda_{\textrm{min}},\lambda_{\textrm{max}}]$ and we used the light implementation of the Precision Lasso. As for HOLP, $\bbeta$ is estimated by
$\hat{\bbeta}_{\textrm{HOLP}}=\bX^T(\bX\bX^T)^{-1}\by$.
Then, for each $s$ in $\{1,\dots, p\}$, the components of $\bbeta$ which are estimated as non null are the $s$ largest among the $|\hat{\bbeta}_{\textrm{HOLP},j}|$,
where $\hat{\bbeta}_{\textrm{HOLP},j}$ denotes the $j$th components of $\hat{\bbeta}_{\textrm{HOLP}}$.
In this case, the parameter controlling the sparsity level of the estimator of $\bbeta$ is $s$. It has a similar role as $\lambda$ in the previous approaches.

\begin{figure}[!h]
  \begin{center}
    \includegraphics[scale=0.5]{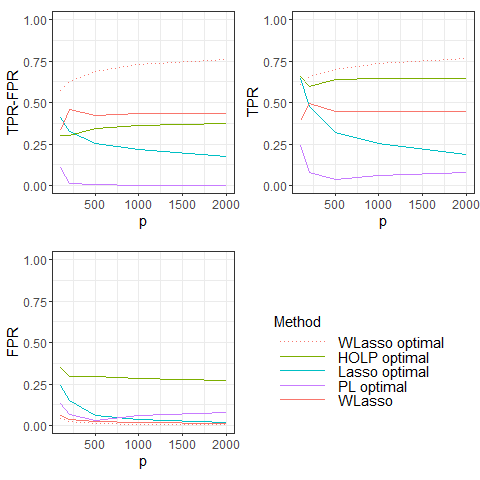}
  \end{center}
  \caption{Top left: max(TPR-FPR) for Lasso, HOLP, Precision Lasso (PL) and (TPR-FPR) for WLasso obtained for the $\lambda$ chosen by the strategy proposed in
    Section \ref{subsec:lambda} (solid line). Results obtained for the optimal choice of $\lambda$ for WLasso (dotted line). Corresponding TPR (top right)
    and FPR (bottom) when $\bSigma$ has the block-wise correlation
  structure defined in (\ref{eq:SPAC}) with parameters $(\alpha_{1} , \alpha_{2} , \alpha_{3})=(0.3, 0.5, 0.7)$, $b=0.5$ and $n=50$.\label{fig:Diff357}}
\end{figure}

\begin{figure}[!h]
  \begin{center}
    \includegraphics[scale=0.5]{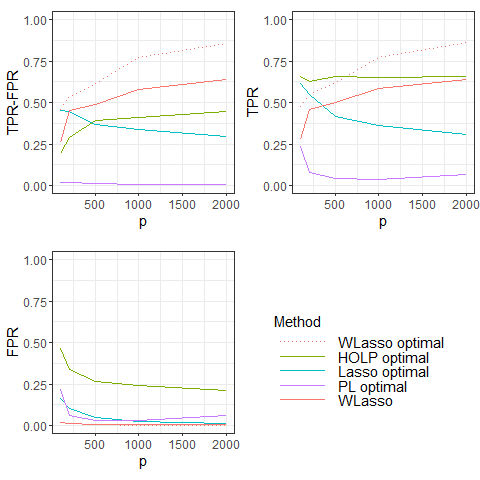}
  \end{center}
\caption{Top left: max(TPR-FPR) for Lasso, HOLP, Precision Lasso (PL) and (TPR-FPR) for WLasso obtained for the $\lambda$ chosen by the strategy proposed in Section \ref{subsec:lambda} (solid line). Results obtained for the optimal choice of $\lambda$ for WLasso (dotted line). Corresponding TPR (top right) and FPR (bottom) when $\bSigma$ has the block-wise correlation
  structure defined in (\ref{eq:SPAC}) with parameters $(\alpha_{1} , \alpha_{2} , \alpha_{3})=(0.5, 0.7, 0.9)$, $b=0.5$ and $n=50$. \label{fig:Diff579}}
\end{figure}

The corresponding results are displayed in Figures \ref{fig:Diff357} and \ref{fig:Diff579} in the case where $n=50$ and $b=0.5$ and
$\bSigma$ has the block-wise correlation
structure defined in (\ref{eq:SPAC}) with parameters $(\alpha_{1} , \alpha_{2} , \alpha_{3})=(0.3, 0.5, 0.7)$ and $(0.5, 0.7, 0.9)$, respectively.
The top left part of these figures displays the largest difference between TPR and FPR for different values of $p$, which corresponds to an optimal choice of the parameters. For WLasso, we also display the results obtained when
the parameter $\lambda$ is chosen by using the strategy proposed in Section \ref{subsec:lambda}, $\gamma=0.95$ and $\bSigma$ is estimated using the procedure explained in Section \ref{sec:estim_Sigma}. The corresponding TPR and FPR for each method are displayed in the top right part and bottom part of the figures, respectively.
Note that we also conducted experiments in the case where $b=1$. Since
the conclusions are very similar, the corresponding figures are given in the Supplementary material.

We can see from Figures \ref{fig:Diff357} and \ref{fig:Diff579} that WLasso outperforms the other methods: the TPR is one of the largest while the FPR is the smallest. HOLP has a larger TPR than WLasso. However, the associated FPR is much larger. It has moreover to be noticed
that Lasso, HOLP and Precision Lasso are favored with respect to WLasso since their parameters were chosen to optimize their performance in terms of TPR
and FPR whereas, in WLasso, the parameter $\lambda$
was chosen by using the strategy of Section \ref{subsec:lambda} and $\bSigma$ was estimated. 
%
%

\begin{figure}[!h]
  \begin{center}
    \includegraphics[scale=0.5]{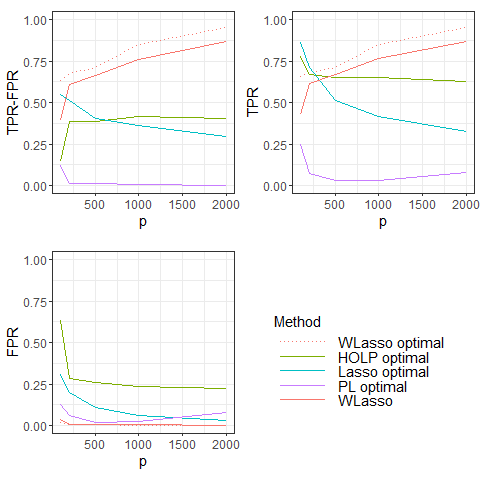}
  \end{center}
  \caption{Top left: max(TPR-FPR) for Lasso, HOLP, Precision Lasso (PL) and (TPR-FPR) for WLasso obtained for the $\lambda$ chosen by the strategy
    proposed in Section \ref{subsec:lambda} (solid line). Results obtained for the optimal choice of $\lambda$ for WLasso (dotted line).
    Corresponding TPR (top right) and FPR (bottom) when $\bSigma$ has the block-wise correlation
  structure defined in (\ref{eq:SPAC}) with parameters $(\alpha_{1} , \alpha_{2} , \alpha_{3})=(0.3, 0.5, 0.7)$, $b=0.5$ and $n=100$.\label{fig:Diff357_100}}
\end{figure}

Figure \ref{fig:Diff357_100} displays the results when the sample size $n$ is increased and equal to 100.  We observe from this figure that the
overall performance has been improved and that our approach outperforms the others especially in the case where $p$ is large.
  Similar results are obtained in the case where $b=1$ and $(\alpha_1, \alpha_2, \alpha_3)$ = $(0.5, 0.7, 0.9)$.
  We refer the reader to the Supplementary material for further details.

\begin{figure}[!h]
  \begin{center}
    \includegraphics[scale=0.5]{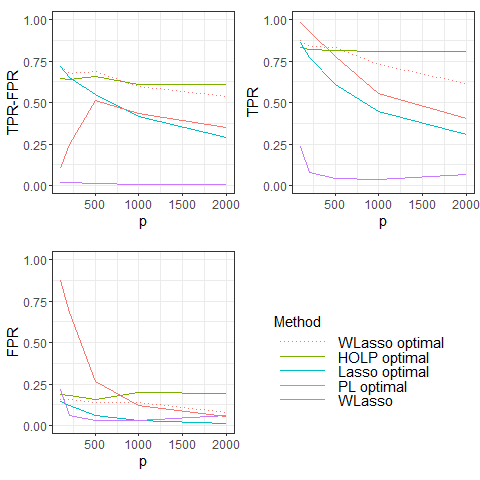}
  \end{center}
\caption{Top left: max(TPR-FPR) for Lasso, HOLP, Precision Lasso (PL) and (TPR-FPR) for WLasso obtained for the $\lambda$ chosen by the strategy proposed in Section \ref{subsec:lambda} (solid line). Results obtained for the optimal choice of $\lambda$ for WLasso (dotted line). Corresponding TPR (top right) and FPR (bottom) when $\bSigma=\textrm{Id}$, $b=0.5$ and $n=50$.\label{fig:Diff000}}
\end{figure}

Figure \ref{fig:Diff000} displays the performance of the different methodologies in the case where $\bSigma=\textrm{Id}$, $n=50$ and $b=0.5$, that
is in the case where there is no correlation between the biomarkers (columns of $\bX$).
We can see from this figure that even in this case, our method, which is
designed for handling the correlation between the biomarkers, obtains similar results as the Lasso except for small values of $p$.
However, when $\lambda$ is chosen to obtain optimal results in terms of the difference between TPR and FPR, our approach achieves the best performance
with HOLP. In the case where $n=100$, our approach obtains the best results, see the Supplementary material.


\subsection{Numerical performance}

\begin{figure}
\centering  \includegraphics[scale=0.25]{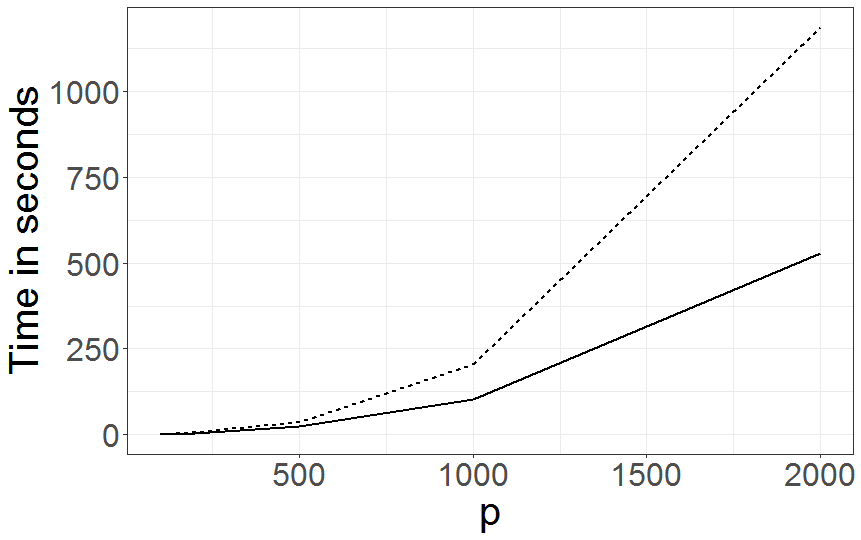}
  \caption{Computational time of our approach WLasso when the parameter ``maxsteps'' has the default value, namely 2000 (dotted line)
    and when maxsteps=500 (solid line).  \label{fig:time_wlasso}}
\end{figure}

Figure \ref{fig:time_wlasso} displays the computational times of our approach implemented in the R package \texttt{WLasso}
for different values of $p$ and of the parameter ``maxsteps'' (maximum number of steps/$\lambda$s considered in the algorithm)
involved in the
\texttt{genlasso} R package.  The timings were obtained on a workstation with
8GB of RAM and Intel Core i5 (2.4GHz) CPU. We can see from this figure that it takes only 6 minutes for processing data with our  approach when $n=50$
and $p=2000$.

\begin{figure}
 \centering  \includegraphics[scale=0.25]{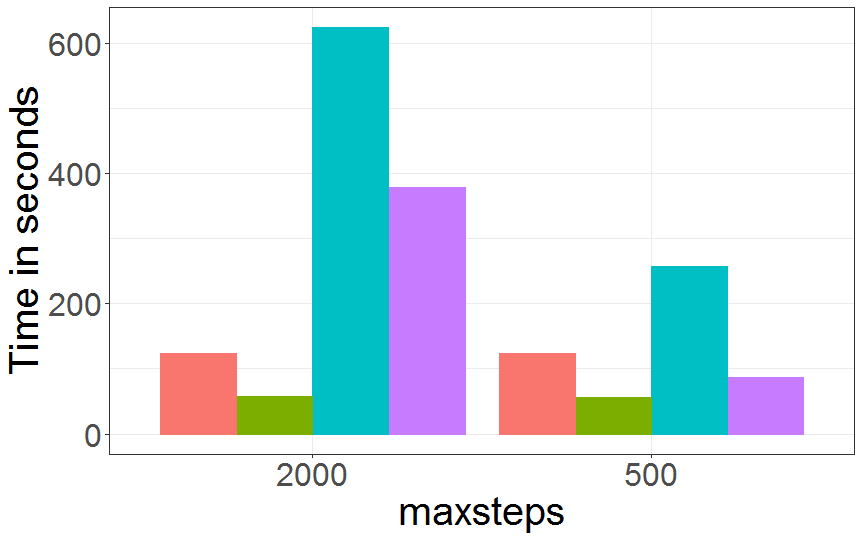}
  \caption{Time allocation for each part of our method WLasso for $p=2000$ and two values of the parameter ``maxsteps''. The time spent in the first three steps
    described in Section \ref{sec:summary} is displayed in the first three bars in red, green, blue, respectively except the correction stage of Steps 3 and 4 (Equations (\ref{eq:beta_tilde_thresh}) and (\ref{eq:beta_hat})) which is given in the fourth bar in purple.\label{fig:time_maxsteps}}
\end{figure}

Moreover, we can observe from Figure \ref{fig:time_maxsteps} that the most time consuming
step of WLasso is the one where the generalized Lasso criterion is used (blue part in Figure \ref{fig:time_maxsteps}).
However, the computational time of this step was divided
by two when the parameter ``maxsteps'' was changed from 2000 (default value) to 500 without changing the variable selection results.

 \section{Application to gene expression data in breast cancer}\label{sec:real}

 We applied the previously detailed methods to publicly available data at Gene Expression Omnibus database (www.ncbi.nlm.nih.gov/geo), with accession code GSE2990, see \cite{Sotiriou:2006}. A total of $n=189$ tumor samples from patients with breast cancer were available and their microarray data have been
collected on 2,283 probes. Expression data has been preprocessed and normalized as in the original publication. A filtering step based on the interquartile range (IQR) was considered to remove some probes as in \cite{Gentleman:2005}.
We removed probes with $\textrm{IQR} < 1.5$ and those which lack of annotation. The remaining $p=1,112$ probes were then standardized.
The goal of the application is to identify genes associated with the development of breast cancer. To this end, the expression of the gene BRCA1 (BReast CAncer gene 1)
was considered as the variable $\by$ to explain (response variable). BRCA1 is a well known human tumor suppressor gene that helps to repair
damaged DNA, thus the mutation of this gene can notably increase the risk of breast cancer, see \cite{Rosen:2003}. The standardized 1,112 probes were considered as explanatory variables and
their correlation is displayed in the Supplementary material.

We compared the performance of the approaches investigated in Section \ref{sec:numexp} in terms of genes selection.
For our approach WLasso, we used the methodology described in Section \ref{sec:estim_Sigma} for 
estimating the correlation matrix $\bSigma$. Assuming that it has the block-wise correlation structure (\ref{eq:SPAC}), we estimated the coefficients $\alpha_1$,
$\alpha_2$ and $\alpha_3$ by $\hat{\alpha}_{1}=0.17, \hat{\alpha}_{2} = 0.21$ and $\hat{\alpha}_{3} = 0.52$, respectively.
As for the Lasso method, the parameter $\lambda$ was chosen by cross-validation and the number of variables to be selected was fixed to 50
for Precision Lasso and HOLP in order to select approximately the same number of variables as with the Lasso.

Table \ref{Tab:application} given in the Supplementary material provides the list of genes corresponding to the selected probes for each method.
Unfortunately, HOLP could not provide any results since it requires the computation of the inverse of the
matrix $\bX\bX^{T}$ which is not invertible in this case. The matrix $\bX^{T}$ is indeed not full rank in this dataset.
The database for annotation, visualization and integrated discovery (DAVID) version 6.8 proposed by \cite{Dennis:2003} was used to highlight
the genes potentially related to the development of breast cancer, which correspond to true positives. 
Based on this tool, 8 genes were considered as true positives for WLasso and Lasso which have only one common gene: GSTT1.
However, Lasso selected more false positives than WLasso (44 vs. 26). Among the 50 variables selected by Precision Lasso,
6 are identified as true positives by DAVID including the well-known ESR1 gene in breast cancer.
Based on this annotation, Precision Lasso detected less true positives than Lasso.
Moreover, Wlasso identified more true positives with less false positives than Precision Lasso. Nevertheless, this application is for illustration
purpose only and not for a fair evaluation or comparison of the methods.


\section{Conclusion}\label{sec:conclusion}



In this paper, we proposed an innovative, efficient and fully data-driven method to deal with the variable selection issue
in high-dimensional frameworks where the active variables are highly correlated with the non-active ones which is implemented in the
\texttt{WLasso} R package available from the CRAN.
The proposed WLasso method has been assessed and compared with other methods in a simulation study with several scenarios. 
In the highly correlated setting, WLasso successfully identifies more true positives with limited false positives as compared with the classical Lasso.
Contrary to HOLP, WLasso still works when several columns are linearly dependent and does not suffer from the inflation of noise introduced by the preconditioning. Compared with the recent Precision Lasso approach, which aims to deal with the same issue, WLasso obtained better results in terms of selection accuracy in the different settings considered.
WLasso is also very computationally
efficient and demonstrated its abilities to properly identify genes related to breast cancer from a publicly available gene expression dataset.
 However, the following directions could be considered to improve its performance.
 
Firstly, the method that we used for estimating $\bSigma$ could be improved by using more sophisticated approaches such as \cite{perrotdocks2018estimation}.
Secondly, our way of choosing the parameter $\lambda$ for the final model selection could also be improved by considering cross-validation or stability selection.
Until now, a simple approach has been considered to avoid computation time and performed quite well especially for moderate to high simple size.
Thirdly, most of the computational time of our approach is spent in the application of the generalized Lasso criterion. Hence, for an application to genomic datasets having more than twenty thousands of variables, it could be worth speeding it up. This will be the subject of future work.




\section*{Funding}

This work was supported by the Association Nationale Recherche Technologie (ANRT).

\bibliographystyle{chicago}


\newpage

\section*{Supplementary material}

  This supplementary material provides additionnal figures and a table for the paper: ``A variable selection approach for highly correlated predictors in high-dimensional genomic data''.

\begin{figure}[!h]
  \begin{center}
    \includegraphics[scale=0.5]{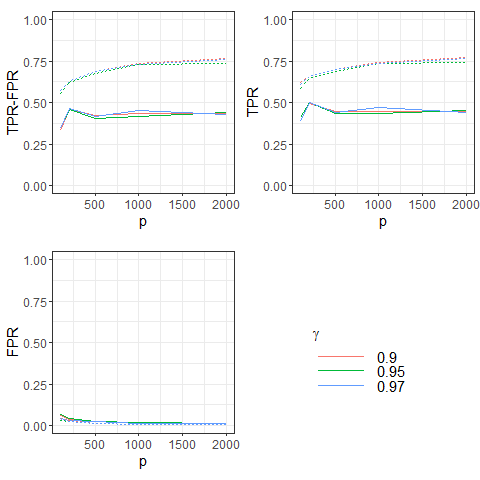}
  \end{center}
  \vspace{-5mm}
  \caption{(TPR-FPR) and the corresponding True Positive Rate and False Positive Rate for  $(\alpha_{1}, \alpha_{2}, \alpha_{3})=(0.3, 0.5, 0.7)$, $b=0.5$
    and $n=50$. Dotted line: optimal choice of $\lambda$, solid line: choice of $\lambda$ explained in Section \ref{subsec:lambda} of the paper.
    \label{fig:Opt_compare_suppl}}
\end{figure}

Figure \ref{fig:Opt_compare_suppl} illustrates Section \ref{subsec:lambda}.

\begin{figure}
  \begin{center}
    \includegraphics[scale=0.5]{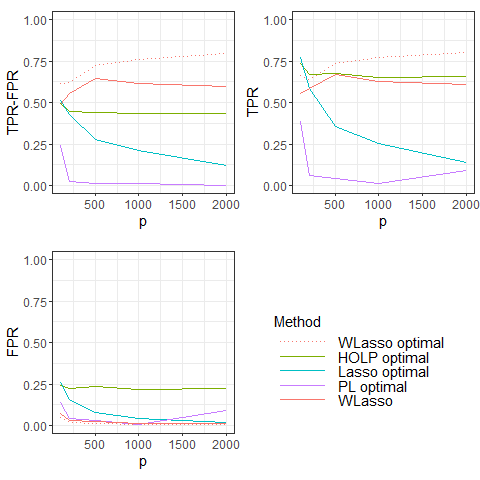}
  \end{center}
  \caption{Top left: max(TPR-FPR) for Lasso, HOLP, Precision Lasso (PL) and (TPR-FPR) for WLasso obtained for the $\lambda$ chosen by the strategy proposed in Section \ref{subsec:lambda}  of the paper (solid line). Results obtained for the optimal choice of $\lambda$ for WLasso (dotted line). Corresponding TPR (top right) and FPR (bottom) when $\bSigma$ has the block-wise correlation structure defined in (\ref{eq:SPAC}) of the paper with parameters $(\alpha_{1},\alpha_{2},\alpha_{3})=(0.3, 0.5, 0.7)$, $b=1$ and $n=50$. \label{fig:357_1_50}}
\end{figure}

\begin{figure}
  \begin{center}
    \includegraphics[scale=0.5]{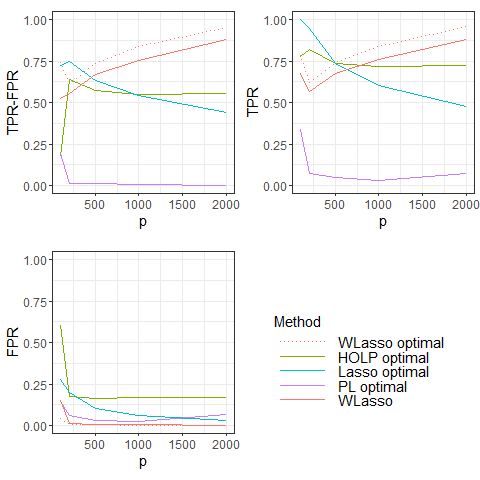}
  \end{center}
\caption{Top left: max(TPR-FPR) for Lasso, HOLP, Precision Lasso (PL) and (TPR-FPR) for WLasso obtained for the $\lambda$ chosen by the strategy proposed in Section \ref{subsec:lambda} (solid line). Results obtained for the optimal choice of $\lambda$ for WLasso (dotted line). Corresponding TPR (top right) and FPR (bottom) when $\bSigma$ has the block-wise correlation
  structure defined in (\ref{eq:SPAC}) with parameters $(\alpha_{1},\alpha_{2},\alpha_{3})=(0.3, 0.5, 0.7)$, $b=1$ and $n=100$. \label{fig:357_1_100}}
\end{figure}

\begin{figure}
  \begin{center}
    \includegraphics[scale=0.5]{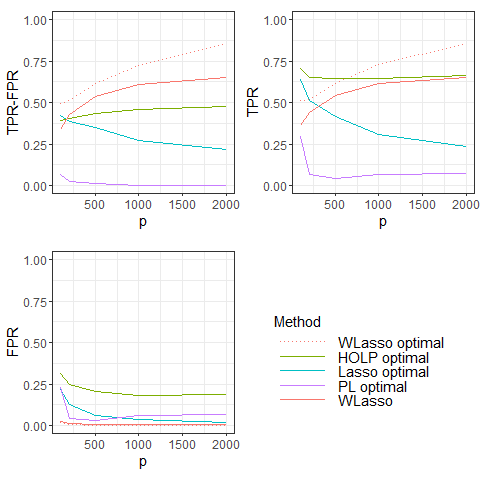}
  \end{center}
\caption{Top left: max(TPR-FPR) for Lasso, HOLP, Precision Lasso (PL) and (TPR-FPR) for WLasso obtained for the $\lambda$ chosen by the strategy proposed in Section \ref{subsec:lambda} (solid line). Results obtained for the optimal choice of $\lambda$ for WLasso (dotted line). Corresponding TPR (top right) and FPR (bottom) when $\bSigma$ has the block-wise correlation
  structure defined in (\ref{eq:SPAC}) with parameters $(\alpha_{1},\alpha_{2},\alpha_{3})=(0.5, 0.7, 0.9)$, $b=1$ and $n=50$. \label{fig:579_1_50}}
\end{figure}

\begin{figure}
  \begin{center}
    \includegraphics[scale=0.5]{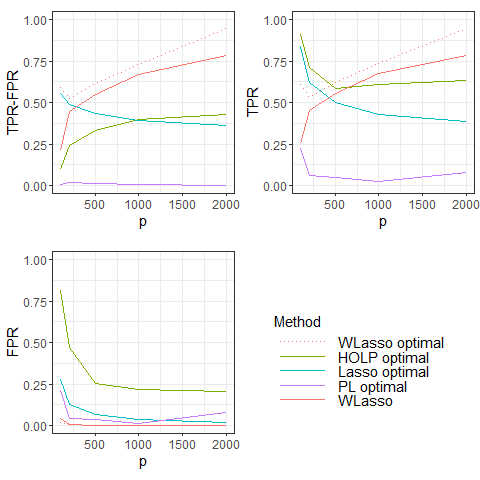}
  \end{center}
\caption{Top left: max(TPR-FPR) for Lasso, HOLP, Precision Lasso (PL) and (TPR-FPR) for WLasso obtained for the $\lambda$ chosen by the strategy proposed in Section \ref{subsec:lambda}  of the paper (solid line). Results obtained for the optimal choice of $\lambda$ for WLasso (dotted line). Corresponding TPR (top right) and FPR (bottom) when $\bSigma$ has the block-wise correlation
  structure defined in (\ref{eq:SPAC})  of the paper with parameters $(\alpha_{1},\alpha_{2},\alpha_{3})=(0.5, 0.7, 0.9)$, $b=0.5$ and $n=100$.\label{fig:579_05_100}}
\end{figure}

\begin{figure}
  \begin{center}
    \includegraphics[scale=0.5]{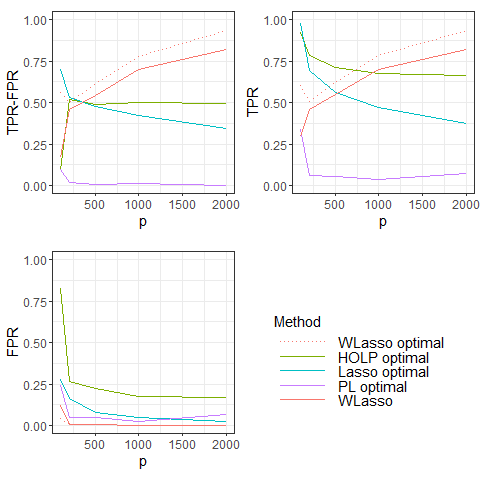}
  \end{center}
\caption{Top left: max(TPR-FPR) for Lasso, HOLP, Precision Lasso (PL) and (TPR-FPR) for WLasso obtained for the $\lambda$ chosen by the strategy proposed in Section \ref{subsec:lambda}  of the paper (solid line). Results obtained for the optimal choice of $\lambda$ for WLasso (dotted line). Corresponding TPR (top right) and FPR (bottom) when $\bSigma$ has the block-wise correlation
  structure defined in (\ref{eq:SPAC})  of the paper with parameters $(\alpha_{1},\alpha_{2},\alpha_{3})=(0.5, 0.7, 0.9)$, $b=1$ and $n=100$. \label{fig:579_1_100}}
\end{figure}

\begin{figure}
  \begin{center}
    \includegraphics[scale=0.5]{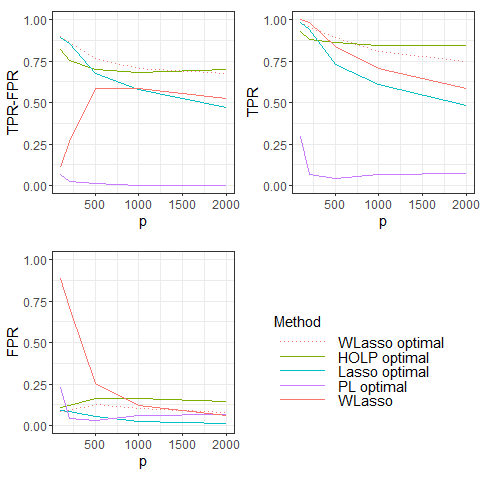}
  \end{center}
\caption{Top left: max(TPR-FPR) for Lasso, HOLP, Precision Lasso (PL) and TPR-FPR for WLasso obtained for the $\lambda$ chosen by the strategy proposed in Section \ref{subsec:lambda}  of the paper (solid line). Results obtained for the optimal choice of $\lambda$ for WLasso (dotted line). Corresponding TPR (top right) and FPR (bottom) when $\bSigma=\textrm{Id}$, $b=1$ and $n=50$. \label{fig:000_1_50}}
\end{figure}

\begin{figure}
  \begin{center}
    \includegraphics[scale=0.5]{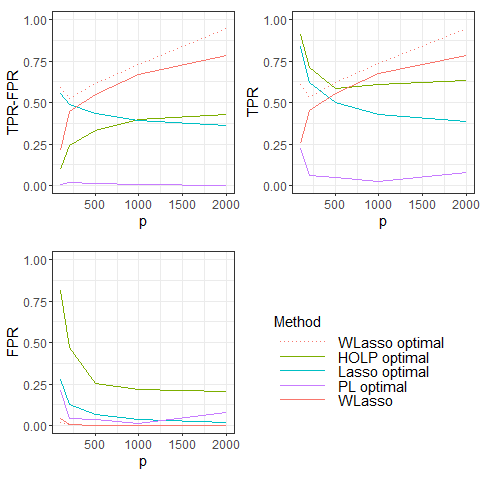}
  \end{center}
\caption{Top left: max(TPR-FPR) for Lasso, HOLP, Precision Lasso (PL) and TPR-FPR for WLasso obtained for the $\lambda$ chosen by the strategy proposed in Section \ref{subsec:lambda}  of the paper (solid line). Results obtained for the optimal choice of $\lambda$ for WLasso (dotted line). Corresponding TPR (top right) and FPR (bottom) when $\bSigma=\textrm{Id}$, $b=0.5$ and $n=100$. \label{fig:000_05_100}}
\end{figure}

\begin{figure}
  \begin{center}
    \includegraphics[scale=0.5]{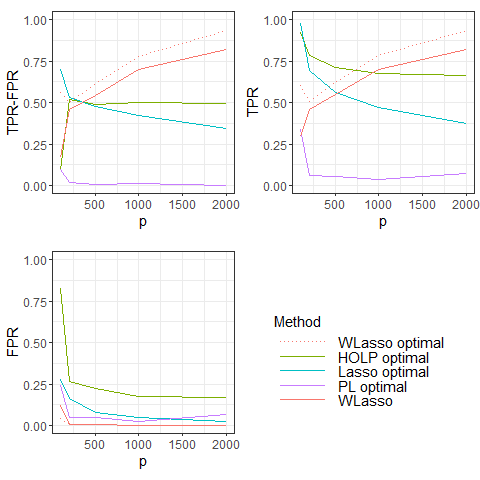}
  \end{center}
\caption{Top left: max(TPR-FPR) for Lasso, HOLP, Precision Lasso (PL) and TPR-FPR for WLasso obtained for the $\lambda$ chosen by the strategy proposed in Section \ref{subsec:lambda}  of the paper (solid line). Results obtained for the optimal choice of $\lambda$ for WLasso (dotted line). Corresponding TPR (top right) and FPR (bottom) when $\bSigma=\textrm{Id}$, $b=1$ and $n=100$. \label{fig:000_1_100}}
\end{figure}

Figures \ref{fig:357_1_50}, \ref{fig:357_1_100},  \ref{fig:579_1_50}, \ref{fig:579_05_100}, \ref{fig:579_1_100},
\ref{fig:000_1_50}, \ref{fig:000_05_100} and \ref{fig:000_1_100} provide similar results as those displayed in Figure \ref{fig:Diff357} of the paper in the following cases:

\begin{itemize}
\item Block-wise correlation structure for $\bSigma$ with $(\alpha_{1},\alpha_{2},\alpha_{3})=(0.3, 0.5, 0.7)$ with
  \begin{itemize}
  \item $b=1$ and $n=50$
  \item $b=1$ and $n=100$
  \end{itemize}
 \item Block-wise correlation structure for $\bSigma$ with $(\alpha_{1},\alpha_{2},\alpha_{3})=(0.5, 0.7,0.9)$ with
  \begin{itemize}
  \item $b=1$ and $n=50$
  \item $b=0.5$ and $n=100$
  \item $b=1$ and $n=100$
  \end{itemize}
\item $\bSigma$ is equal to identity with
  \begin{itemize}
  \item $b=1$ and $n=50$
  \item $b=0.5$ and $n=100$
  \item $b=1$ and $n=100$
  \end{itemize}
\end{itemize}

\begin{figure}[!htbp]
\centering
  \includegraphics[scale=0.5]{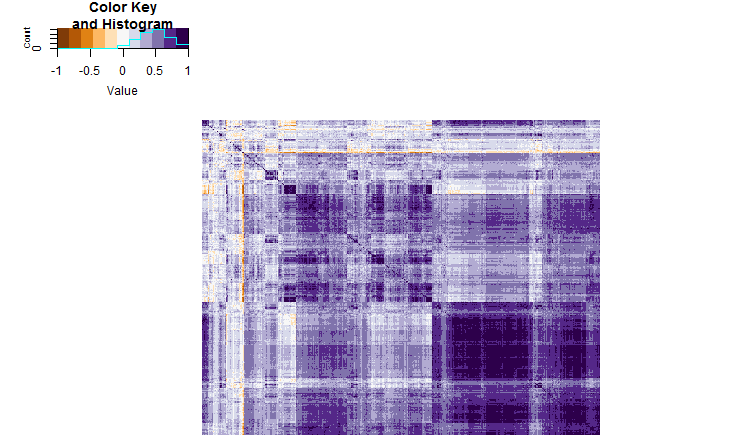}
\caption{Heatmap of correlations between the probes.\label{fig:App_hm}}
\end{figure}

Figures \ref{fig:App_hm} and Table \ref{Tab:application} give additionnal information for the Application Section.

\begin{table}[]
  \centering
    \caption{Selected genes from each method, genes related to breast cancer are in bold. \label{Tab:application}}
{\scriptsize{
    \begin{tabular}{|c|c|}
  \hline
  & selected genes \\\hline
WLasso &     \makecell{ \textbf{TOP2A}, NPEPPS, INSIG1, \textbf{PLK2}, TRIM2,\\ \textbf{CYP1B1}, KIF5C, ATXN1, \textbf{PTTG1}, NUCB2,\\ SEMA3C, \textbf{GSTT1}, LGALS4, \textbf{KIF11}, \textbf{NEK2},\\ \textbf{RAD51}, DNALI1, AGFG2, IAPP, IFI16,\\ CITED2, SLC19A2, WSB1, RET, FRMD4B,\\ MYBL1, SELENBP1, CALU, ZNF451, IGHM,\\ AZGP1, NCAPG, ANO1, ZNF226}  \\\hline
  Lasso  &\makecell{CCL5, WSB1, MAPKAPK2, FHL1, TNC,\\ ALCAM, ATP6V1A, RIOK3, RGS2, \textbf{USP1}, \\TRIM14, \textbf{LPL}, GGH, \textbf{GSTT1}, SMN1,\\ MMP11, PSMB9, DST, \textbf{RAD54L}, TFAP2A,\\ DSC2, GABRP, INPP4B, \textbf{GHR}, CD36,\\ PTP4A3, ASS1, H2BFS, ANXA3, \textbf{CXCL12},\\ \textbf{TPX2}, CACNA1D, GOLGA8A, ATG5, SC5D,\\ PTN, C1R, WWP1, DPT, \textbf{MUC1},\\ LRRC15, SMA4, CCZ1B, ASNSD1, COPS4,\\ PSD3, VAV3, MS4A4A, KLF13, QPCTL,\\ PLA2G12A, MUM1}     \\\hline
  PL     & \makecell{ H2AFZ, BTG2, SDC1,\textbf{IGF2R}, INSIG1,\\ ALCAM, SDHD, ACADM, FBLN1, SNRPE,\\ N4BP2L2 , SLC25A37, \textbf{IGF1R}, PPFIBP1, SYT1,\\ SORBS2, AGA, IFI44L, TFF3, DSC2,\\ MYB,\textbf{ESR1}, SETBP1, \textbf{FGFR2}, GOLGA8A,\\ DNAJB6 , CD24,\textbf{HLA-DQB1}, JUP, SULF1,\\ RDX, COL14A1, NFIB, COL6A1, KRT6B,\\ NBPF10, N4BP2L2, DGKH, RGS1, ZSCAN18,\\ LZTFL1, \textbf{GREM1}, C1orf115, DUSP12, KLHL2,\\ ARMC9, DENND1B, TMPRSS3, WIZ, NTRK2  }   \\\hline
  HOLP   &      \\\hline
    \end{tabular}
    }}
\end{table}

\end{document}